\documentstyle[preprint,aps,prbbib,graphicx]{revtex}
  
\begin{document}
\thispagestyle{empty}
\begin{center}{COSMOLOGICAL RELATIVITY: A GENERAL-RELATIVISTIC THEORY FOR THE
ACCELERATING EXPANDING UNIVERSE$^\star$}\end{center}
\vspace{1cm}
\begin{center}{Moshe Carmeli and Silvia Behar}\end{center} 
\begin{center}{Department of Physics, Ben Gurion University, Beer Sheva 84105, 
Israel}\end{center}
\begin{center}{(E-mail: carmelim@bgumail.bgu.ac.il \hspace{1cm} 
silviab@bgumail.bgu.ac.il)}\end{center}
\begin{center}{ABSTRACT}\end{center}
Recent observations of distant supernovae imply, in defiance of expectations,
that the universe growth is accelerating, contrary to what has always been
assumed that the expansion is slowing down due to gravity. In this paper a
general-relativistic cosmological theory that gives a direct relationship
between distances and redshifts in an expanding universe is presented. The 
theory is actually a generalization of Hubble's law taking gravity into 
account by means of Einstein's theory of general relativity. The theory 
predicts that the universe can have three phases of expansion, decelerating,
constant and accelerating, but it is shown that at present the first two cases
are excluded, although in the past it had experienced them. Our theory shows 
that the universe now is definitely in a stage of accelerating expansion,
confirming the recent experimental results.
\vspace{2cm}\newline
$^\star$Paper dedicated to Professor Sir Hermann Bondi on the occasion of his 
80th birthday. 
\newpage
The general-relativistic theory of cosmology started in 1922 with the
remarkable work of A. Friedmann [1,2], who solved the Einstein gravitational
field equations and found that they admit non-static cosmological solutions
presenting an expandeing universe. Einstein, believing that the universe 
should be static and unchanged forever, suggested a modification to his
gravitational field equations by adding to them the so-called cosmological
term which can stop the expansion.

Soon after that E. Hubble [3,4] found experimentally that the far-away galaxies 
are receding from us, and that the farther the galaxy the bigger its velocity.
In simple words, the universe is indeed expanding according to a simple 
physical law that gives the relationship between the receding velocity and
the distance,
$$\mbox{\bf v}=H_0\mbox{\bf R}.\eqno(1)$$ 
Equation (1) is usually referred to as the Hubble law, and $H_0$ is called
the Hubble constant. It is tacitly assumed that the velocity is proportional
to the actual measurement of the redshift $z$ of the receding objects by using
the non-relativistic relation $z=\mbox{\bf v}/c$, where $c$ is the speed of 
light in vacuum. 

The Hubble law does not resemble standard dynamical physical laws that are 
familiar in physics. Rather, it is a {\it cosmological equation of state} of
the kind one has in thermodynamics that relates the pressure, volume and 
temperature, $pV=RT$ [5]. It is this Hubble's equation of state that will be
extended so as to include gravity by use of the full Einstein theory of 
general relativity. The obtained results will be very simple, expressing 
distances in terms of redshifts; depending on the value of $\Omega=\rho/\rho_c$
we will have accelerating, constant and decelerating expansions, corresponding 
to $\Omega<1$, $\Omega=1$ and $\Omega>1$, respectively. But the last two cases
will be shown to be excluded on physical evidence, although the universe had
decelerating and constant expansions before it reached its present 
accelerating expansion stage. As is well known, the standard theory does not
deal with this problem.

Before presenting our theory, and in order to fix the notation, we very 
briefly review the existing theory [6,7]. In the four-dimensional curved 
space-time describing the universe, our spatial three-dimensional space is
assumed to be isotropic and homogeneous. Co-moving coordinates, in which 
$g_{00}=1$ and $g_{0k}=0$, are employed [8,9]. Here, and throughout our paper,
low-case Latin indices take the values 1, 2, 3 whereas Greek indices will
take the values 0, 1, 2, 3, and the signature will be $(+ - - -)$. The 
four-dimensional space-time is split into $1\bigoplus 3$ parts, and the 
line-element is subsequently written as $ds^2=dt^2-dl^2$, where $dl^2={}^{(3)}
g_{kl}dx^k dx^l=-g_{kl}dx^k dx^l$, and the $3\times 3$ tensor ${}^{(3)}g_{kl}
\equiv -g_{kl}$ describes the geometry of the three-dimensional space at a 
given instant of time. In the above equations the speed of light $c$ was taken 
as unity.

Because of the isotropy and homogeneity of the three-geometry, it follows that
the curvature tensor must have the form
$${}^{(3)}R_{mnsk}=K\left[{}^{(3)}g_{ms}{}^{(3)}g_{nk}-{}^{(3)}g_{mk}
{}^{(3)}g_{ns}\right],\eqno(2)$$   
where $K$ is a constant, the curvature of the three-dimensional space, which 
is related to the Ricci scalar by ${}^{(3)}R=-6K$ [10]. By simple geometrical
arguments one then finds that
$$dl^2=\left(1-r^2/R^2\right)^{-1}dr^2+r^2\left(d\theta^2+\sin^2\theta d\phi^2
\right),\eqno(3)$$
where $r<R$. Furthermore, the curvature tensor corresponding to the metric (3)
satisfies Eq. (2) with $K=1/R^2$. In the ``spherical" coordinates $\left(t,r,
\theta,\phi\right)$ we thus have
$$g_{11}=-\left(1-r^2/R^2\right)^{-1}.\eqno(4)$$ 
$R$ is called the radius of curvature of the universe (or the expansion 
parameter) and its value is determined by the Einstein field equations.

One then has three cases: (1) a universe with positive curvature for which
$K=1/R^2$; (2) a universe with negative curvature, $K=-1/R^2$; and (3) a 
universe with zero curvature, $K=0$. The $g_{11}$ component for the 
negative-curvature universe is given by
$$g_{11}=-\left(1+r^2/R^2\right)^{-1}.\eqno(5)$$
For the zero-curvature universe one lets $R\rightarrow\infty$.

Although general relativity theory asserts that all coordinate systems are
equally valid, in this theory one has to change variables in order to get 
the ``right"
solutions of the Einstein field equations according to the type of the 
universe. Accordingly, one makes the substitution $r=R\sin\chi$ for the
positive-curvature universe, and $r=R\sinh\chi$ for the negative-curvature
universe. Not only that, but the time-like coordinate is also changed into 
another one $\eta$ by the transformation $dt=Rd\eta$. The corresponding line
elements then become:
$$ds^2=R^2\left(\eta\right)\left[d\eta^2-d\chi^2-\sin^2\chi\left(d\theta^2
+\sin^2\theta d\phi^2\right)\right]\eqno(6a)$$
for the positive-curvature universe,
$$ds^2=R^2\left(\eta\right)\left[d\eta^2-d\chi^2-\sinh^2\chi\left(d\theta^2
+\sin^2\theta d\phi^2\right)\right]\eqno(6b)$$
for the negative-curvature universe, and
$$ds^2=R^2\left(\eta\right)\left[d\eta^2-dr^2-r^2\left(d\theta^2
+\sin^2\theta d\phi^2\right)\right]\eqno(6c)$$
for the flat three-dimensional universe. In the sequel, we will see that the
time-like coordinate in our theory will take one more different form. 

The Einstein field equations are then employed in order to determine the 
expansion parameter $R(\eta)$. In fact only one field equation is needed,
$$R_0^0-\frac{1}{2}\delta_0^0R+\Lambda\delta_0^0=8\pi GT_0^0,\eqno(7)$$
where $\Lambda$ is the cosmological constant, and $c$ was taken as unity. In 
the Friedmann models one takes $\Lambda=0$, and in the comoving coordinates 
used one easily finds that $T_0^0=\rho$, the mass density. While this choice
of the energy-momentum tensor is acceptable in standard general relativity and
in Newtonian gravity, we will argue in the sequel that it is not so for 
cosmology. At any rate, using $\rho\left(t\right)=M/2\pi^2R^3$, where $M$ is
the ``mass" and $2\pi^2R^3$ is the ``volume" of the universe, one obtains
$$3\left[\left(dR/dt\right)^2+1\right]/R^2=4GM/\pi R^3+\Lambda,\eqno(8a)$$
or, in terms of $\eta$ along with taking $\Lambda=0$,
$$3\left[\left(dR/d\eta\right)^2+R^2\right]/R=4GM/\pi,\eqno(9a)$$ 
The solution of this equation is
$$R={}^\star R\left(1-\cos\eta\right),\eqno(10a)$$
where ${}^\star R$ is a constant, ${}^\star R=2GM/3\pi$, and from $dt=R
\mbox{\hspace{2pt}}d\eta$
we obtain
$$t={}^\star R\left(\eta-\sin\eta\right).\eqno(11a)$$
Equations (10a) and (11a) are those of a cycloid, and give a full 
representation for the radius of curvature of the universe.

Similarly, one obtains for the negative-curvature universe the analog to Eqs.
(8a) and (9a),
$$3\left[\left(dR/dt\right)^2-1\right]/R^2=4GM/\pi R^3+\Lambda,\eqno(8b)$$
$$3\left[\left(dR/d\eta\right)^2-R^2\right]/R=4GM/\pi,\eqno(9b)$$
the solution of which is given by
$$R={}^\star R\left(\cosh\eta-\eta\right),\eqno(10b)$$
$$t={}^\star R\left(\sinh\eta-\eta\right).\eqno(11b)$$ 

Finally, for the universe with a flat three-dimensional space the Einstein
field equations yield the analog to Eqs. (8a) and (9a), 
$$3\left(dR/dt\right)^2/R^2=4GM/\pi R^3+\Lambda,\eqno(8c)$$  
$$3\left(dR/d\eta\right)^2/R=4GM/\pi.\eqno(9c)$$ 
As a function of $t$, the solution is
$$R=\left(3GM/\pi\right)^{1/3}t^{2/3}.\eqno(10c)$$

An extension of the Friedmann models was carried out by Lema\^\i tre, who
considered universes with zero energy-momentum but with a non-zero 
cosmological constant. One then again obtains positive-curvature, 
negative-curvature and zero-curvature (also called de Sitter) universes. 
While these models are of interest mathematically they have little, if any,
relation to the physical universe. 

In the final analysis, it follows that the expansion of the universe is 
determined by the so-called cosmological parameters. These are the mass
density $\rho$, the Hubble constant $H$ and the deceleration parameter $q$.
We will not go through this, however, since we will concentrate on the theory
with dynamical variables that are actually measured by astronomers: distances,
redshifts and the mass density.

One of the Friedmann theory assumptions is that the type of the universe is 
determined
by $\Omega=\rho/\rho_c$, where $\rho_c=3H_0^2/8\pi G$, which requires that the 
sign of $(\Omega-1)$ must not change throughout the evolution of the universe
so as to change the kind of the universe from one to another. That means in
this theory, the universe has only one kind of curvature throughout its
evolution and does not permit going from one curvature into another. In other 
words the universe has been and will be in only one 
form of expansion. It is not 
obvious, however, that this is indeed a valid assumption whether theoretically
or experimentally. As will be shown in the sequel, the universe has actually
three phases of expansion and it {\it does} go from one to the second and
then to the third phase. 

A new outlook at the universe expansion can be achieved and is presented here.
The new theory has the following features: (1) the dynamical variables are
Hubble's, i.e. distances and redshifts, the actually-measured quantities by
astronomers; (2) it is fully general relativistic; (3) it includes two
universal constants, the speed of light in vacuum $c$, and the Hubble time in 
the absence of gravity $\tau$ (might also be called the {\it Hubble time in
vacuum}); (4) the redshift parameter $z$ is taken as the time-like
coordinate; (5) the energy-momentum tensor is represented differently; and
(6) it predicts that the universe has three phases of expansion: decelerating,
constant and accelerating, but it is now in the stage of accelerating 
expansion phase after passing the other two phases.

Our starting point is Hubble's cosmological equation of state, Eq. (1). One
can keep the velocity $v$ in equation (1) or replace it with the redshift 
parameter $z$ by means of $z=v/c$. Since {\bf R}$=(x_1,x_2,x_3)$, the square 
of Eq. (1) then yields
$$c^2H_0^{-2}z^2-\left(x_1^2+x_2^2+x_3^2\right)=0.\eqno(12)$$
Our aim is to write our equations in an invariant way so as to enable us to
extend them to curved space. Equation (12) is not invariant since $H_0^{-1}$
is the Hubble time at present. At the limit of zero gravity, Eq. (12) will
have the form
$$c^2\tau^2z^2-\left(x_1^2+x_2^2+x_3^2\right)=0,\eqno(13)$$ 
where $\tau$ is Hubble's time in vacuum, which is a {\it universal constant}
the numerical value of which will be determined in the sequel by relating it 
to $H_0^{-1}$ at different situations. Equation (13) provides the basis of a
cosmological special relativity and has been investigated extensively [11-16].

In order to make Eq. (13) adaptable to curved space we write it in a 
differential form:
$$c^2\tau^2dz^2-\left(dx_1^2+dx_2^2+dx_3^2\right)=0,\eqno(14)$$              
or, in covariant form in flat space,
$$ds^2=\eta_{\mu\nu}dx^\mu dx^\nu=0,\eqno(15a)$$
where $\eta_{\mu\nu}$ is the ordinary Minkowskian metric, and our coordinates
are $(x^0,x^1,x^2,x^3)=(c\tau z,x_1,x_2,x_3)$. Equation (15a) expresses the
null condition, familiar from light propagation in space, but here it 
expresses the universe expansion in space. The generalization of Eq. (15a) to 
a covariant form in curved space can immediately be made by replacing the 
Minkowskian metric $\eta_{\mu\nu}$ by the curved Riemannian geometrical 
metric $g_{\mu\nu}$,
$$ds^2=g_{\mu\nu}dx^\mu dx^\nu=0,\eqno(15b)$$  
obtained from solving the Einstein field equations.

Because of the spherical symmetry nature of the universe, the metric we seek
is of the form [8]
$$ds^2=c^2\tau^2dz^2-e^\lambda dr^2-r^2\left(d\theta^2+\sin^2\theta d\phi^2
\right),\eqno(16)$$
where co-moving coordinates, as in the Friedmann theory, are used and 
$\lambda$ is a function of the radial distance $r$. The metric (16) is static
and solves the Einstein field equation (7). When looking for static solutions, 
Eq. (7) can also be written as
$$e^{-\lambda}\left(\lambda'/r-1/r^2\right)+1/r^2=8\pi GT_0^0 \eqno(17)$$
when $\Lambda$ is taken zero, and where a prime denotes differentiation with
respect to $r$.

In general relativity theory one takes for $T_0^0=\rho$. So is the situation 
in Newtonian gravity where one has the Poisson equation $\nabla^2\phi=4\pi G
\rho$. At points where $\rho=0$ one solves the vacuum Einstein field equations 
and the Laplace equation $\nabla^2\phi=0$ in Newtonian gravity. In both 
theories a null (zero) solution is allowed as a trivial case. In cosmology, 
however, there exists no situation at which $\rho$ can be zero because the
universe is filled with matter. In order to be able to have zero on the 
right-hand side of Eq. (17) we take $T_0^0$ not as equal to $\rho$ but to
$\rho-\rho_c$, where $\rho_c$ is chosen by us now as a {\it constant} given
by $\rho_c=3/8\pi G\tau^2$. This 
approach has been presented and used in earlier work [17].   

The solution of Eq. (17), with $T_0^0=\rho-\rho_c$, is given by 
$$e^{-\lambda}=1-\left(\Omega-1\right)r^2/c^2\tau^2,\eqno(18)$$
where $\Omega=\rho/\rho_c$. Accordingly, if $\Omega>1$ we have $g_{rr}=
-\left(1-r^2/R^2\right)^{-1}$, where
$$R^2=c^2\tau^2/\left(\Omega-1\right),\eqno(19a)$$
exactly equals to $g_{11}$ given by Eq. (4) for the positive-curvature
Friedmann universe that is obtained in the standard theory by purely 
geometrical manipulations. If $\Omega<1$, we can write $g_{rr}=       
-\left(1+r^2/R^2\right)^{-1}$ with
$$R^2=c^2\tau^2/\left(1-\Omega\right),\eqno(19b)$$
which is equal to $g_{11}$ given by Eq. (5) for the negative-curvature
Friedmann universe. In the above equations $r<R$.

Moreover, we know that the Einstein field equations for these cases are given
by Eqs. (8) which, in our new notation, have the form
$$\left[\left(dR/dz\right)^2+c^2\tau^2\right]/R^2=\left(\Omega-1\right),
\eqno(20a)$$  
$$\left[\left(dR/dz\right)^2-c^2\tau^2\right]/R^2=\left(\Omega-1\right).
\eqno(20b)$$
As is seen from these equations, if one neglects the first term in the square
brackets with respect to the second ones, $R^2$ will be exactly reduced to
their values given by Eqs. (19).

The expansion of the universe can now be determined from the null condition
$ds=0$, Eq. (15b), using the metric (16). Since the expansion is radial, one
has $d\theta=d\phi=0$, and the equation obtained is
$$dr/dz=c\tau\left[1+\left(1-\Omega\right)r^2/c^2\tau^2\right]^{1/2}.
\eqno(21)$$

The second term in the square brackets of Eq. (21) represents the deviation
from constant expansion due to gravity. For without this term, Eq. (21)
reduces to $dr/dz=c\tau$ or $dr/dv=\tau$, thus $r=\tau v+$constant. The 
constant can be taken zero if one assumes, as usual, that at $r=0$ the 
velocity should also vanish. Accordingly we have $r=\tau v$ or $v=\tau^{-1}r$.
When $\Omega=1$, namely when $\rho=\rho_c$, we have a constant expansion.
 
The equation of motion (21) can be integrated exactly by the substitutions
$$\begin{array}{cc}\sin\chi=\alpha r/c\tau;&\Omega>1,\\\end{array}\eqno(22a)$$
$$\begin{array}{cc}\sinh\chi=\beta r/c\tau;&\Omega<1,\\\end{array}\eqno(22b)$$
where 
$$\begin{array}{cc}\alpha=\left(\Omega-1\right)^{1/2},&
\beta=\left(1-\Omega\right)^{1/2}.\\\end{array}\eqno(23)$$

For the $\Omega>1$ case a straightforward calculation, using Eq. (22a), gives
$$dr=\left(c\tau/\alpha\right)\cos\chi d\chi,\eqno(24)$$
and the equation of the universe expansion (21) yields
$$d\chi=\alpha dz.\eqno(25a)$$
The integration of this equation gives
$$\chi=\alpha z+\mbox{\rm constant}.\eqno(26a)$$
The constant can be determined, using Eq. (22a). For at $\chi=0$, we have 
$r=0$ and $z=0$, thus
$$\chi=\alpha z,\eqno(27a)$$
or, in terms of the distance, using (22a) again,
$$\begin{array}{cc}r\left(z\right)=\left(c\tau/\alpha\right)\sin\alpha z;&
\alpha=\left(\Omega-1\right)^{1/2}.\\\end{array}\eqno(28a)$$
This is obviously a decelerating expansion.

For $\Omega<1$, using Eq. (22b), then a similar calculation yields for the
universe expansion (21)
$$d\chi=\beta dz,\eqno(25b)$$
thus
$$\chi=\beta z+\mbox{\rm constant}.\eqno(26b)$$ 
Using the same initial conditions used above then give
$$\chi=\beta z,\eqno(27b)$$  
and in terms of distances,
$$\begin{array}{cc}r\left(z\right)=\left(c\tau/\beta\right)\sinh\beta z;&
\beta=\left(1-\Omega\right)^{1/2}.\\\end{array}\eqno(28b)$$
This is now an accelerating expansion.

For $\Omega=1$ we have, from Eq. (21), 
$$d^2r/dz^2=0.\eqno(25c)$$
The solution is, of course,
$$r\left(z\right)=c\tau z.\eqno(28c)$$
This is a constant expansion.

It will be noted that the last solution can also be obtained directly from
the previous two ones for $\Omega>1$ and $\Omega<1$ by going to the limit
$z\rightarrow 0$, using L'Hospital lemma, showing that our solutions are
consistent. It will be shown later on that the constant expansion is just a
transition stage between the decelerating and the accelerating expansions as 
the universe evolves toward its present situation. 

Figure 1 describes the Hubble diagram of the above solutions for the three
types of expansion for values of $\Omega$ from 100 to 0.24. The figure 
describes the three-phase evolution of the universe. Curves (1) to (5) 
represent the stages of  decelerating expansion according to Eq. (28a). As
the density of matter $\rho$ decreases, so does $\Omega$ also, the universe
goes over from the lower curves to the upper ones, and it does not have
enough time to close up to a big crunch. The universe, subsequently, goes to
curve (6) with $\Omega=1$, at which time it has a constant expansion for a
fraction of a second. This then followed by going to the upper curves (7) and
(8) with $\Omega<1$ where the universe expands with acceleration according to
Eq. (28b). A curve of this kind fits the present situation of the universe.

In order to decide which of the three cases is the appropriate one at the
present time, we have to write the solutions (28) in the ordinary Hubble law
form $v=H_0r$. To this end we change variables from the redshift parameter $z$
to the velocity $v$ by means of $z=v/c$ for $v$ much smaller than $c$. For
higher velocities this relation is not accurate and one has to use a Lorentz 
transformation in order to relate $z$ to $v$. A simple calculation then shows
that, for receding objects, one has the relations
$$z=\left[\left(1+v/c\right)/\left(1-v/c\right)\right]^{1/2}-1,\eqno(29a)$$ 
$$v/c=z\left(z+2\right)/\left(z^2+2z+2\right).\eqno(29b)$$
We will assume that $v\ll c$ and consequently Eqs. (28) have the forms
$$r\left(v\right)=\left(c\tau/\alpha\right)\sin\left(\alpha v/c\right),
\eqno(30a)$$
$$r\left(v\right)=\left(c\tau/\beta\right)\sinh\left(\beta v/c\right),
\eqno(30b)$$
$$r\left(v\right)=\tau v.\eqno(30c)$$

Expanding now Eqs. (30a) and (30b) and keeping the appropriate terms, then 
yields
$$r=\tau v\left(1-\alpha^2v^2/6c^2\right),\eqno(31a)$$
for the $\Omega>1$ case, and
$$r=\tau v\left(1+\beta^2v^2/6c^2\right),\eqno(31b)$$  
for $\Omega<1$. Using now the expressions for $\alpha$ and $\beta$, given by 
Eq. (23), in Eqs. (31) then both of the latter can be reduced into a single 
equation
$$r=\tau v\left[1+\left(1-\Omega\right)v^2/6c^2\right].\eqno(32)$$
Inverting now this equation by writing it in the form $v=H_0r$, we obtain in 
the lowest approximation for $H_0$ the following:
$$H_0=h\left[1-\left(1-\Omega\right)v^2/6c^2\right],\eqno(33)$$
where $h=\tau^{-1}$. Using $v\approx r/\tau$, or $z=v/c$, we also obtain
$$H_0=h\left[1-\left(1-\Omega\right)r^2/6c^2\tau^2\right]
=h\left[1-\left(1-\Omega\right)z^2/6\right].\eqno(34)$$

Consequently, $H_0$ depends on the distance, or equivalently, on the redshift.
As is seen, $H_0$ has meaning only for $r\rightarrow 0$ or $z\rightarrow 0$,
namely when measured {\it locally}, in which case it becomes $h$.

 In recent years observers have argued for values of $H_0$ as low as 50 and
as high as 90 km/sec-Mpc, some of the recent ones show $80\pm 17$ km/sec-Mpc
[18-26]. There are the so-called ``short" and ``long" distance scales, with 
the higher and the lower values for $H_0$ respectively [27]. Indications are 
that the longer the distance of measurement the smaller the value of $H_0$.
By Eqs. (33) and (34) this is possible only for the case in which $\Omega<1$,
namely when the universe is at an accelerating expansion.

Figures 2 and 3 show the Hubble diagrams of the predicted by theory 
distance-redshift relationship for the accelerating expanding universe at 
present time, whereas figures 4 and 5 show the experimental results [28-29].

Our estimate for $h$, based on published data, is $h\approx 85-90$ km/sec-Mpc.
Assuming $\tau^{-1}\approx 85$ km/sec-Mpc, Eq. (34) then gives
$$H_0=h\left[1-1.3\times 10^{-4}\left(1-\Omega\right)r^2\right],\eqno(35)$$
where $r$ is in Mpc. A computer best-fit can then fix both $h$ and $\Omega$.

In summary, we have presented a general-relativistic theory of cosmology, 
the dynamical variables of which are those of Hubble's, i.e. distances and 
redshifts. The theory describes the universe as having a three-phase evolution
with a decelerating expansion, followed by a constant and an accelerating 
expansion, and it predicts that the universe is now in the latter phase. As 
the density of matter decreases, while the universe is at the decelerating 
phase, it does not have enough time to close up to a big crunch. Rather, it 
goes to the constant expansion phase, and then to the accelerating stage. The
equations obtained for the universe expansion are very simple.

Finally, it is worth mentioning that if one assumes that the matter density at
the present time is such that $\Omega_0=0.24$, then it can be shown that the
time at which the universe had gone over from a decelerating to an 
accelerating expansion, i.e. the constant expansion phase, occured at cosmic 
time $0.03\tau$.

The idea to express cosmological theory in terms of directly-measurable
quantities, such as distances and redshifts, was partially inspired by
Albert Einstein's favourite remarks on the theory of thermodynamics in his
Autobiographical Notes [30].
\begin{center}{ACKNOWLEDGEMENTS}\end{center}
It is a great pleasure to thank Professor Sir Hermann Bondi for his continuous
interest, comments and stimulating suggestions dealing with the contents of
this paper.
We also wish to thank Professor Michael Gedalin for his help with the
diagrams.

\newpage
\begin{center}{FIGURE CAPTIONS}\end{center}     
Fig. 1 Hubble's diagram describing the three-phase evolution of the universe
according to Einstein's general relativity theory. Curves (1) to (5) represent
the stages of {\it decelerating} expansion according to $r(z)=(c\tau/\alpha)
\sin\alpha z$, where $\alpha=(\Omega-1)^{-1/2}$, $\Omega=\rho/\rho_c$, with
$\rho_c$ a {\it constant}, $\rho_c=3/8\pi G\tau^2$, and $c$ and $\tau$ are 
the speed of light and the Hubble time in vacuum (both universal constants).
As the density of matter $\rho$ decreases, the universe goes over from the 
lower curves to the upper ones, but it does not have enough time to close up to
a big crunch. The universe subsequently goes to curve (6) with $\Omega=1$, at
which time it has a {\it constant} expansion for a fraction of a second. This
then followed by going to the upper curves (7)--(8) with $\Omega<1$ where the
universe expands with {\it acceleration} according to $r(z)=(c\tau/\beta)
\sinh\beta z$, where $\beta=(1-\Omega)^{-1/2}$. One of these last curves fits
the present situation of the universe.\vspace{5mm}\newline
Fig. 2 Hubble's diagram of the universe at the present phase of evolution with
accelerating expansion.\vspace{5mm}\newline
Fig. 3 Hubble's diagram describing decelerating, constant and accelerating 
expansions in a logarithmic scale.\vspace{5mm}\newline
Fig. 4 Distance vs. redshift diagram showing the deviation from a constant
toward an accelerating expansion. [Sourse: A. Riess {\it et al., Astron. J.}
{\bf 116}, 1009 (1998)].\vspace{5mm}\newline
Fig. 5 Relative intensity of light and relative distance vs. redshift.
[Sourse: A. Riess {\it et al., Astron. J.} {\bf 116}, 1009 (1998)].  

\newpage
\begin{figure}
\centering
\includegraphics{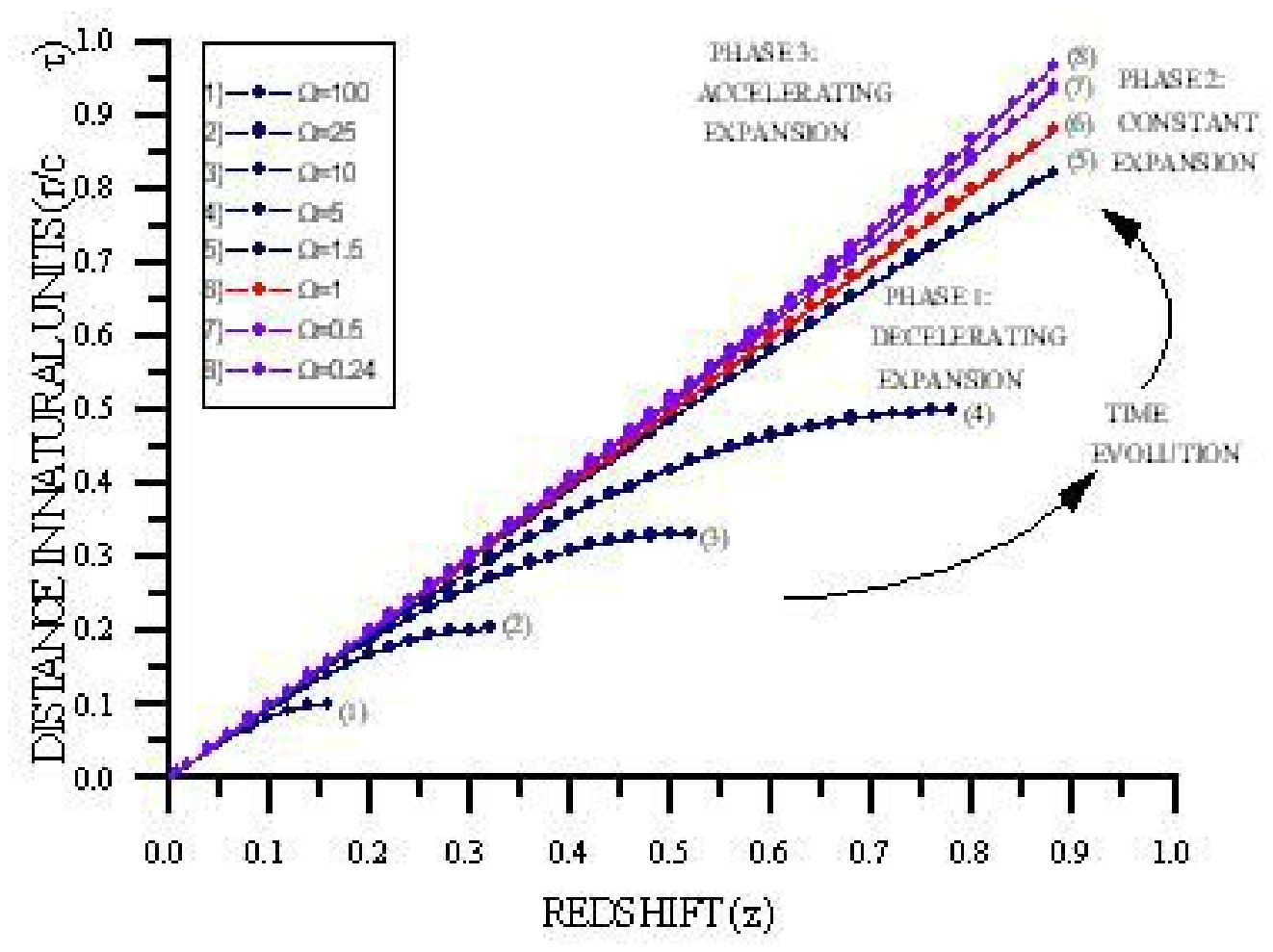}
\caption{}
\end{figure}

\newpage
\begin{figure}
\centering
\includegraphics{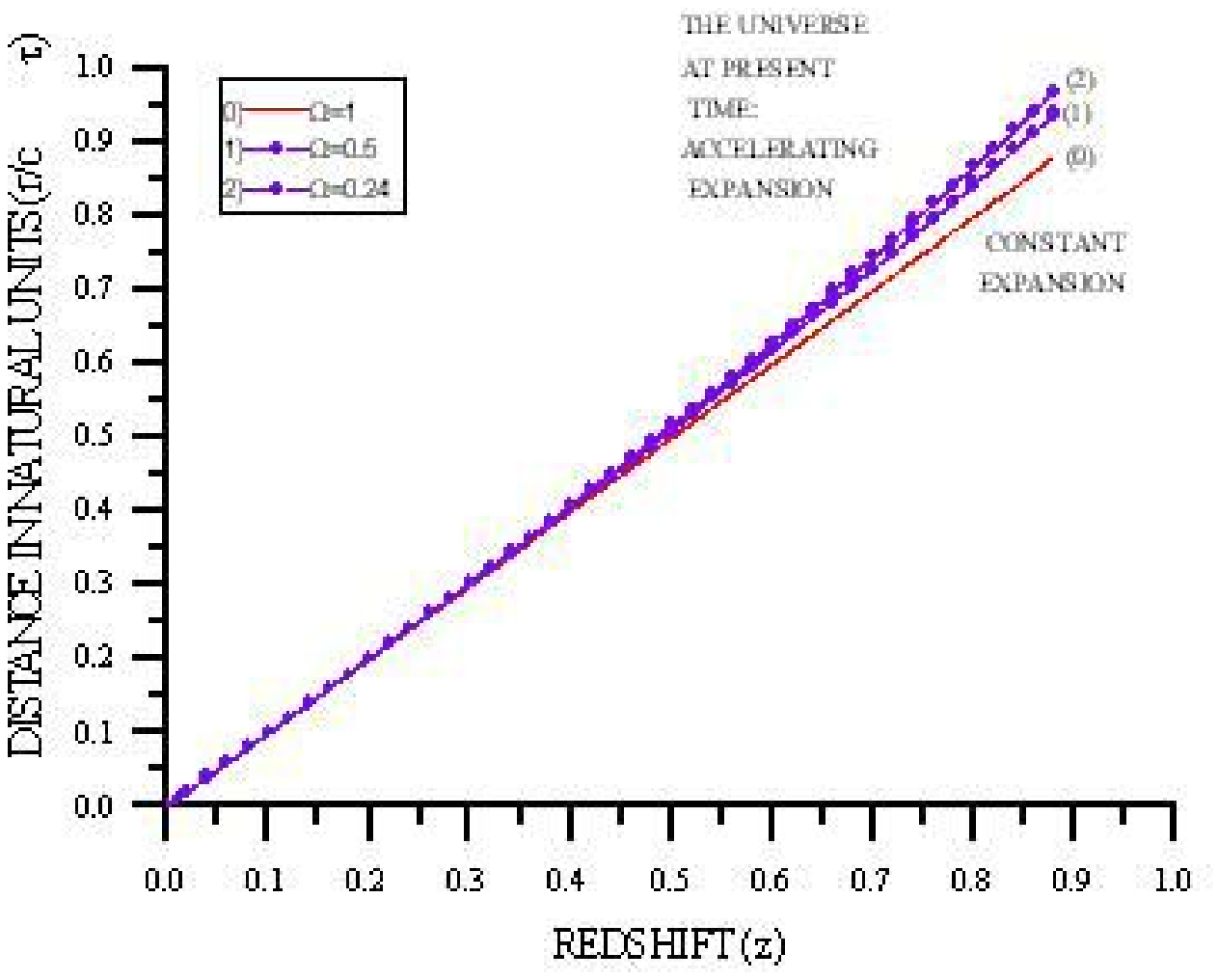}
\caption{}
\end{figure}

\newpage
\begin{figure}
\centering
\includegraphics{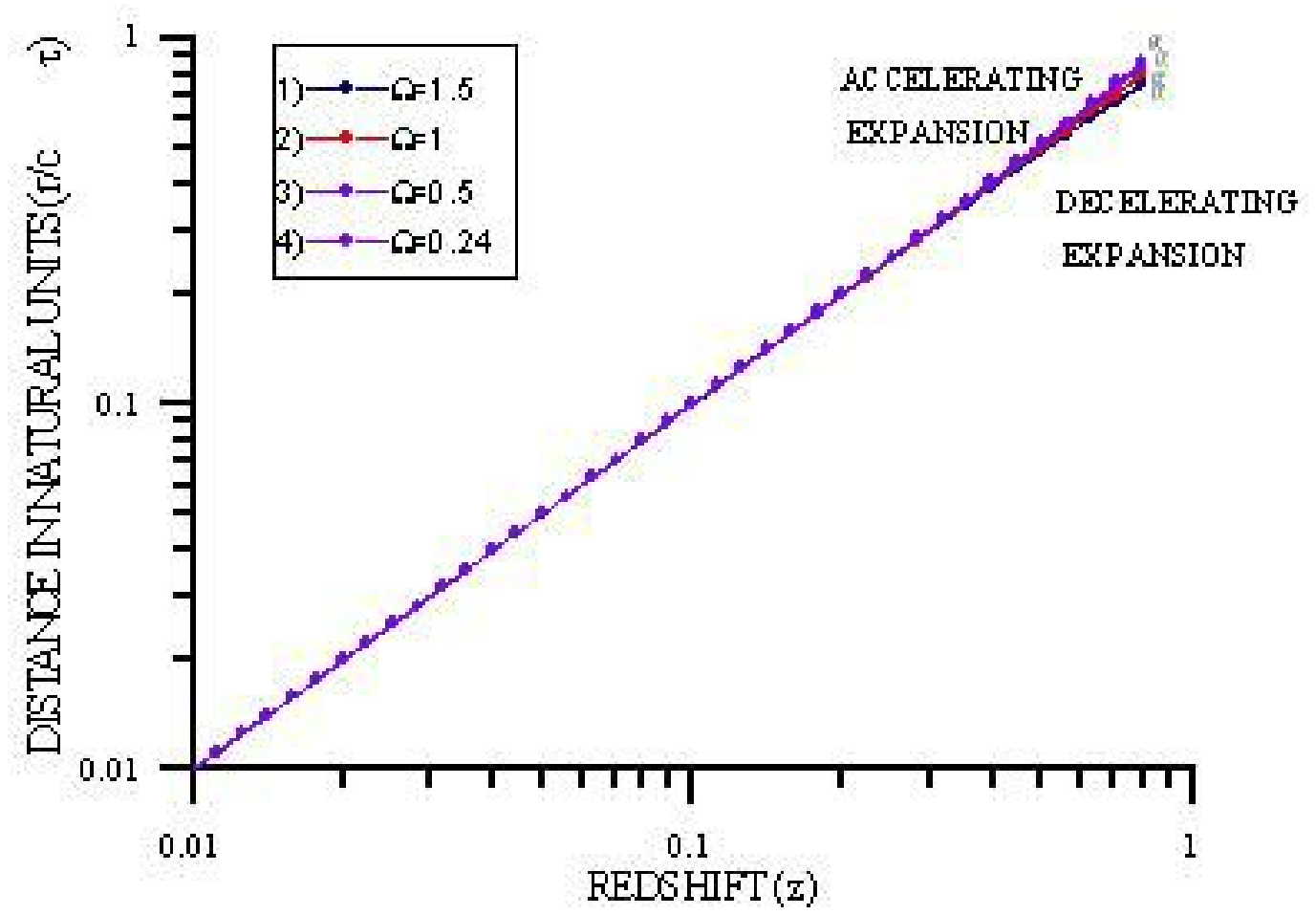}
\caption{}
\end{figure}

\newpage
\begin{figure}
\centering
\includegraphics{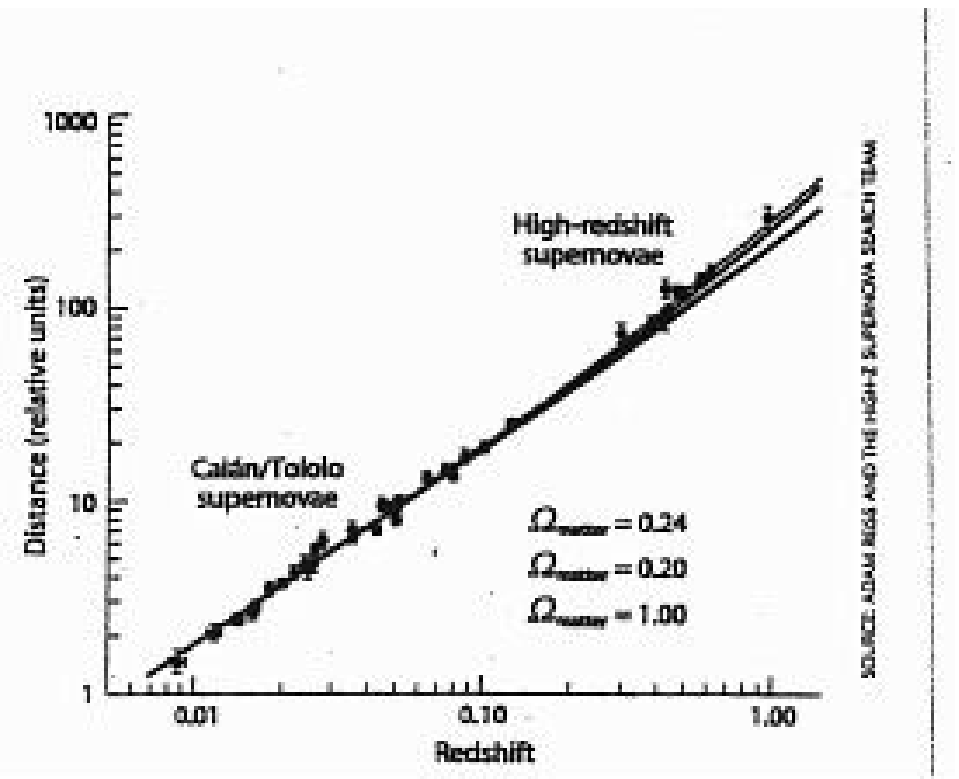}
\caption{}
\end{figure}

\newpage
\begin{figure}
\centering
\includegraphics{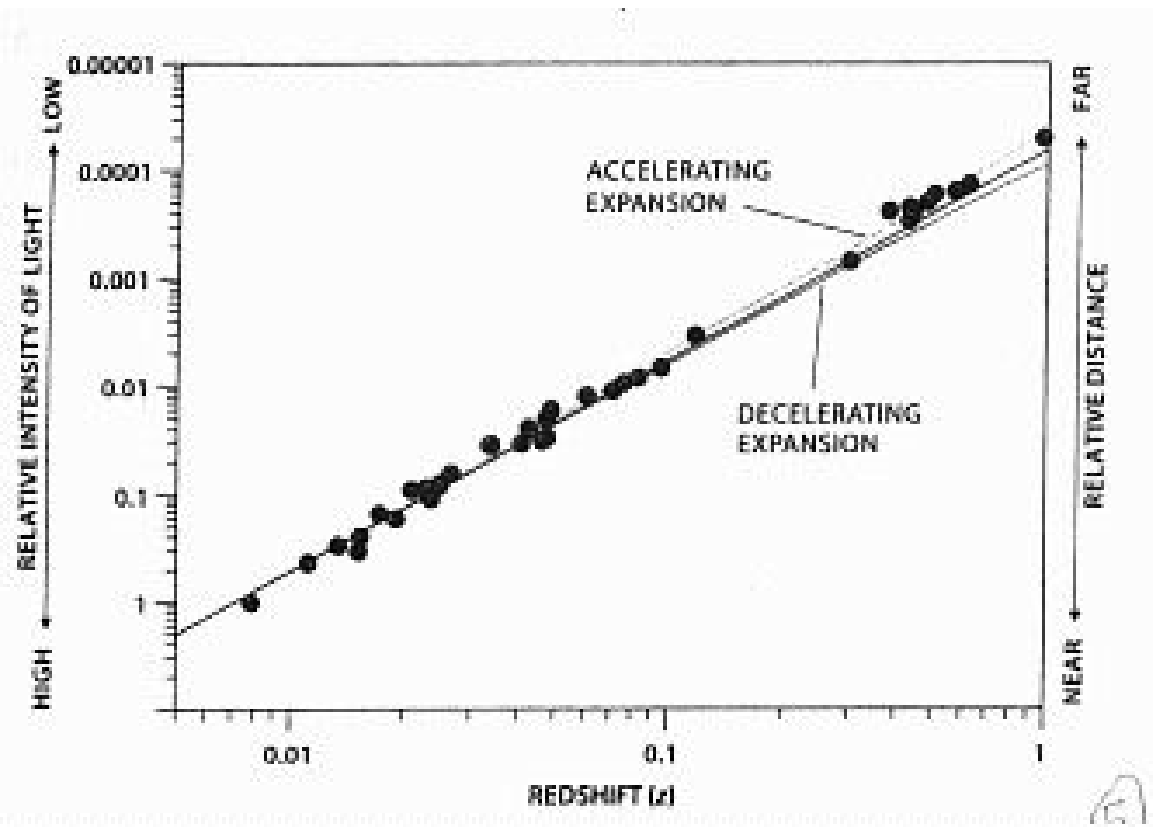}
\caption{}
\end{figure}

\end{document}